\pdfoutput=1
\documentclass[conference,a4paper,final]{IEEEtran}[2015/08/26]

\usepackage{upquote}

\usepackage[ngerman,main=english]{babel}
\addto\extrasenglish{\languageshorthands{ngerman}\useshorthands{"}}

\usepackage[hyphens]{url}

\makeatletter
\g@addto@macro{\UrlBreaks}{\UrlOrds}
\makeatother

\usepackage[zerostyle=b,scaled=.75]{newtxtt}

\usepackage[T1]{fontenc}

\usepackage[
  babel=true, % Enable language-specific kerning. Take language-settings from the languge of the current document (see Section 6 of microtype.pdf)
  expansion=alltext,
  protrusion=alltext-nott, % Ensure that at listings, there is no change at the margin of the listing
  final % Always enable microtype, even if in draft mode. This helps finding bad boxes quickly.
]{microtype}

\DisableLigatures{encoding = T1, family = tt* }

\usepackage{graphicx}

\usepackage{diagbox}

\usepackage{xcolor}

\usepackage{listings}

\definecolor{eclipseStrings}{RGB}{42,0.0,255}
\definecolor{eclipseKeywords}{RGB}{127,0,85}
\colorlet{numb}{magenta!60!black}

\lstdefinelanguage{json}{
    basicstyle=\normalfont\ttfamily,
    commentstyle=\color{eclipseStrings}, % style of comment
    stringstyle=\color{eclipseKeywords}, % style of strings
    numbers=left,
    numberstyle=\scriptsize,
    stepnumber=1,
    numbersep=8pt,
    showstringspaces=false,
    breaklines=true,
    frame=lines,
    string=[s]{"}{"},
    comment=[l]{:\ "},
    morecomment=[l]{:"},
    literate=
        *{0}{{{\color{numb}0}}}{1}
         {1}{{{\color{numb}1}}}{1}
         {2}{{{\color{numb}2}}}{1}
         {3}{{{\color{numb}3}}}{1}
         {4}{{{\color{numb}4}}}{1}
         {5}{{{\color{numb}5}}}{1}
         {6}{{{\color{numb}6}}}{1}
         {7}{{{\color{numb}7}}}{1}
         {8}{{{\color{numb}8}}}{1}
         {9}{{{\color{numb}9}}}{1}
}

\usepackage[autostyle=true]{csquotes}

\defineshorthand{"`}{\openautoquote}
\defineshorthand{"'}{\closeautoquote}

\usepackage{booktabs}

\usepackage{paralist}

\usepackage[%
  square,        % for square brackets
  comma,         % use commas as separators
  numbers,       % for numerical citations;
  sort&compress % as sort but in addition multiple numerical citations
]{natbib}

\usepackage{etoolbox}
\makeatletter
\patchcmd{\NAT@test}{\else \NAT@nm}{\else \NAT@hyper@{\NAT@nm}}{}{}
\makeatother

\usepackage{pdfcomment}

\usepackage{stfloats}\fnbelowfloat

\usepackage[group-minimum-digits=4,per-mode=fraction]{siunitx}
\addto\extrasgerman{\sisetup{locale = DE}}

\usepackage{hyperref}

\hypersetup{
  hidelinks,
  colorlinks=true,
  allcolors=black,
  pdfstartview=Fit,
  breaklinks=true
}

\usepackage[all]{hypcap}

\usepackage[caption=false,font=footnotesize]{subfig}

\usepackage[capitalise,nameinlink,noabbrev]{cleveref}

\crefname{listing}{Listing}{Listings}
\Crefname{listing}{Listing}{Listings}
\crefname{lstlisting}{Listing}{Listings}
\Crefname{lstlisting}{Listing}{Listings}

\usepackage{lipsum}

\usepackage[math]{blindtext}
\usepackage{mwe}
\usepackage[realmainfile]{currfile}
\usepackage{tcolorbox}
\tcbuselibrary{listings}

\DeclareFontFamily{U}{MnSymbolC}{}
\DeclareSymbolFont{MnSyC}{U}{MnSymbolC}{m}{n}
\DeclareFontShape{U}{MnSymbolC}{m}{n}{
  <-6>    MnSymbolC5
  <6-7>   MnSymbolC6
  <7-8>   MnSymbolC7
  <8-9>   MnSymbolC8
  <9-10>  MnSymbolC9
  <10-12> MnSymbolC10
  <12->   MnSymbolC12%
}{}
\DeclareMathSymbol{\powerset}{\mathord}{MnSyC}{180}

\usepackage{xspace}
\makeatletter
\newcommand{\hydash}{\penalty\@M-\hskip\z@skip}
\makeatother

\usepackage{siunitx}

\input glyphtounicode
\begin{document}
% Enable following command if you need to typeset "IEEEpubid".
% See https://bytefreaks.net/tag/ieeeoverridecommandlockouts for details.
%\IEEEoverridecommandlockouts

\title{IoT Droplocks: Wireless fingerprint theft using hacked smart locks}

\author{%
  \IEEEauthorblockN{Steve Kerrison}
  \IEEEauthorblockA{James Cook University, Singapore\\
    steve.kerrison@jcu.edu.au}
}

\IEEEspecialpapernotice{Accepted to appear in 2022 IEEE International Conference on Internet of Things (iThings)\cite{KerrisonDroplocksiThings2022}}

\maketitle

% In case you want to add a copyright statement.
% Works only in the compsoc conference mode.
%
% Source: https://tex.stackexchange.com/a/325013/9075
%
% All possible solutions:
%  - https://tex.stackexchange.com/a/325013/9075
%  - https://tex.stackexchange.com/a/279134/9075
%  - https://tex.stackexchange.com/q/279789/9075 (TikZ)
%  - https://tex.stackexchange.com/a/200330/9075 - for non-compsocc papers
\iftrue
   \IEEEoverridecommandlockouts
   \IEEEpubid{\begin{minipage}{\textwidth}\ \\[40pt] \centering
    Published version DOI: \href{https://doi.org/10.1109/iThings-GreenCom-CPSCom-SmartData-Cybermatics55523.2022.00054}{10.1109/iThings-GreenCom-CPSCom-SmartData-Cybermatics55523.2022.00054}, \textcopyright 2022 IEEE.
     \end{minipage}}
\fi

\def\droplock*{{\em droplock}}

\begin{abstract}
Electronic locks can provide security- and convenience-enhancing features, with fingerprint readers an increasingly common feature in these products. When equipped with a wireless radio, they become a smart lock and join the billions of IoT devices proliferating our world. However, such capabilities can also be used to transform smart locks into fingerprint harvesters that compromise an individual's security without their knowledge. We have named this the \droplock* attack. This paper demonstrates how the harvesting technique works, shows that off-the-shelf smart locks can be invisibly modified to perform such attacks, discusses the implications for smart device design and usage, and calls for better manufacturer and public treatment of this issue.
\end{abstract}

% \fi
\IEEEpeerreviewmaketitle

\section{Introduction}
\label{sec:introduction}
Biometric data such as fingerprints are now widely used as an authentication factor. They secure smart phones, banking activities, staff clock in/out and doors to buildings or rooms. As a ``something you are'' factor, they are valued for their uniqueness and permanent attachment to an individual. However, unlike other factors such as passwords, tokens and keys, there is a limit to how many times they can be changed. Fingerprint-based security measures becomes undesirable if too many compromises occur, and existing technology is not always able to distinguish fake or copied prints.

Despite this, a growing number of devices have built-in fingerprint sensors, most notably smartphones. While there are demonstrations of attacks that bypass or otherwise fool smartphone fingerprint sensors~\cite{MccarthySandy2019SwfG,AroraShefali2019FSDt}, less attention has been given towards using such sensors as a means to steal a fingerprint. This is largely because the sensors and the associated fingerprint processing is handled in a secure domain such as a Trusted Execution Environment~\cite{AndroidBiometricSpecs} and the fingerprint image is not easy to extract from them.

Beyond smartphones, various lower-cost IoT devices also have fingerprint sensors, such as smart locks. These devices generally feature less powerful processors, cheaper sensors and do not provide the same level of security as a smartphone. This is usually deemed acceptable based on the value of the product itself, or what the sensor is meant to protect.

Even though such IoT devices might not directly protect a high-value asset such as a person's bank account, they utilise a shared authentication factor, thus posing a risk to such assets if the biometric data can be harvested. Additionally, these devices may be small, portable, battery powered and have WiFi, Bluetooth or other wireless communication capabilities. This makes them cheap and convenient to use as a lure for unwitting individuals who may divulge their biometric data to such a device and have that data wirelessly collected by a nearby yet out-of-sight adversary. This is akin to the {\em dropped USB} or {\em dropped memory card} attack, whereby malware infiltrates a system at the behest of the user, whose own intrigue and lack of awareness leads them to delivering a payload from an unknown storage device onto a vulnerable computer. In acknowledgement of this heritage, we term our attack the \emph{IoT droplock} or simply \droplock*.

With this premise, we construct the following thesis:

\begin{itemize}
    \item Cheap IoT components can be used to capture and wirelessly harvest fingerprint data.
    \item Commercial off-the-Shelf (COTS) smart locks can be modified to achieve the same goal due to lack of adequate defences in the hardware and software design.
    \item Fingerprints harvested through the above means can then be used for authentication with other systems.
\end{itemize}

\begin{figure}
  \centering
  \includegraphics[angle=90,width=.67\linewidth,trim={2.5cm 3cm 5cm 3cm},clip]{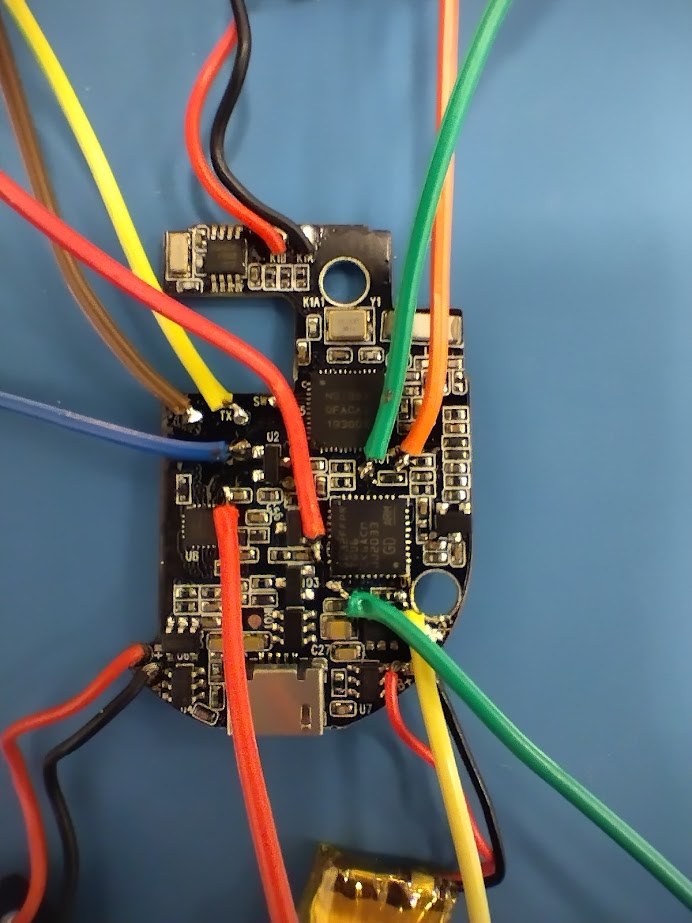}
  \caption{A COTS smart lock prepared to receive a \droplock* firmware.}
  \label{fig:lock-pcb}
\end{figure}

\noindent In this paper, we demonstrate how both self-made and commercial IoT devices (such as that shown in \cref{fig:lock-pcb}) can be used as wireless fingerprint harvesters, explore the implications of this with the support of prior literature, make recommendations for device manufacturers to improve the trustworthiness of their products and propose future work to better understand and influence user behaviour when confronted with potential threats of this nature.

\subsection{Structure}
\label{sec:structure}

\Cref{sec:comppoc} is a demonstration of a self-made fingerprint harvester. \Cref{sec:hacking} then elevates this concept to a full-scale hack of a commercial off-the-shelf (COTS) smart padlock with integrated fingerprint reader and Bluetooth Low Energy (BLE), which can ultimately be performed without device disassembly. Having established the ability to harvest prints, \cref{sec:relatedwork} reviews literature that explores how people are susceptible to ``drop'' style attacks, how fingerprint images can be used fool security systems and what other vulnerabilities exist in electronic and smart locks. We make recommendations in \cref{sec:recommendations}, with a list of security measures that would mitigate \droplock* attack attempts and propose further research, including a behavioural study of how individuals react to \droplock*s and similar devices. Finally, \cref{sec:conclusion} summarises and concludes this paper.

\section{Proof of concept}
\label{sec:comppoc}

The objective of a \droplock* attack is to get a human to supply a fingerprint image to a device that they find (for example placed on the floor), then for that device to wirelessly transmit the image elsewhere for future malicious use. In support of this concept, a simple demonstrator was created. \Cref{tab:poc} shows the major components of the PoC. These could be substituted for other similar parts, provided the micro-controller possesses some form of wireless radio support and the fingerprint reader includes the ability to upload fingerprint images.

\begin{figure}
  \centering
  \includegraphics[width=.9\linewidth]{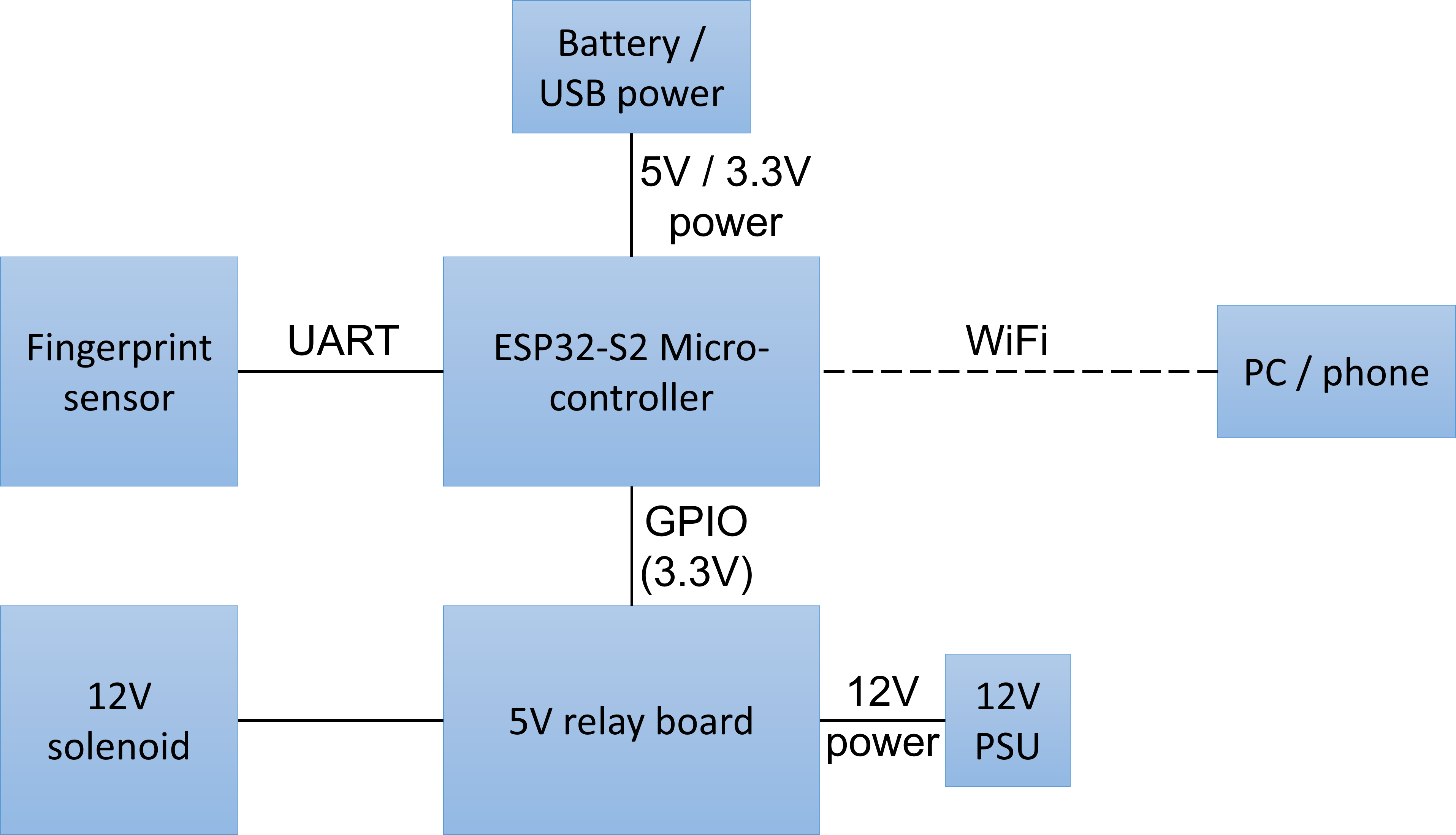}
  \caption{Diagram of the \droplock* PoC component construction.}
  \label{fig:pocdiagram}
\end{figure}

A diagram of the PoC's construction is shown in \cref{fig:pocdiagram}. It relies on an external power supply to provide \SI{12}{\volt} to actuate the relay, mainly due to component availability at the time of construction. Additionally, the ESP32-S2's GPIOs operate at \SI{3.3}{\volt} and the relay board uses \SI{5}{\volt} logic levels, but its active-low signalling design means no additional hardware is required to control the relay.

\begin{figure*}
  \centering
  \begin{tabular}{lp{0.67\textwidth}}
    \toprule
    Item & Purpose \\
    \midrule
    ESP32-S2 board w/LCD & Micro-controller with WiFi, GPIO for relay control, UART for fingerprint sensor, wireless comms for data capture and LCD for visual prompt.     \\
    12V DC solenoid     & Equivalent to locking mechanism present on smart lock.     \\
    5V relay breakout board       & Interface between I/O voltage of micro-controller and actuation voltage of solenoid. \\
    Fingerprint sensor w/UART & Collect fingerprint image and upload to micro-controller. \\
    \bottomrule
  \end{tabular}
  \caption{Components used in \droplock* PoC}
  \label{tab:poc}
\end{figure*}

\begin{figure}
  \centering
  \includegraphics[width=.9\linewidth]{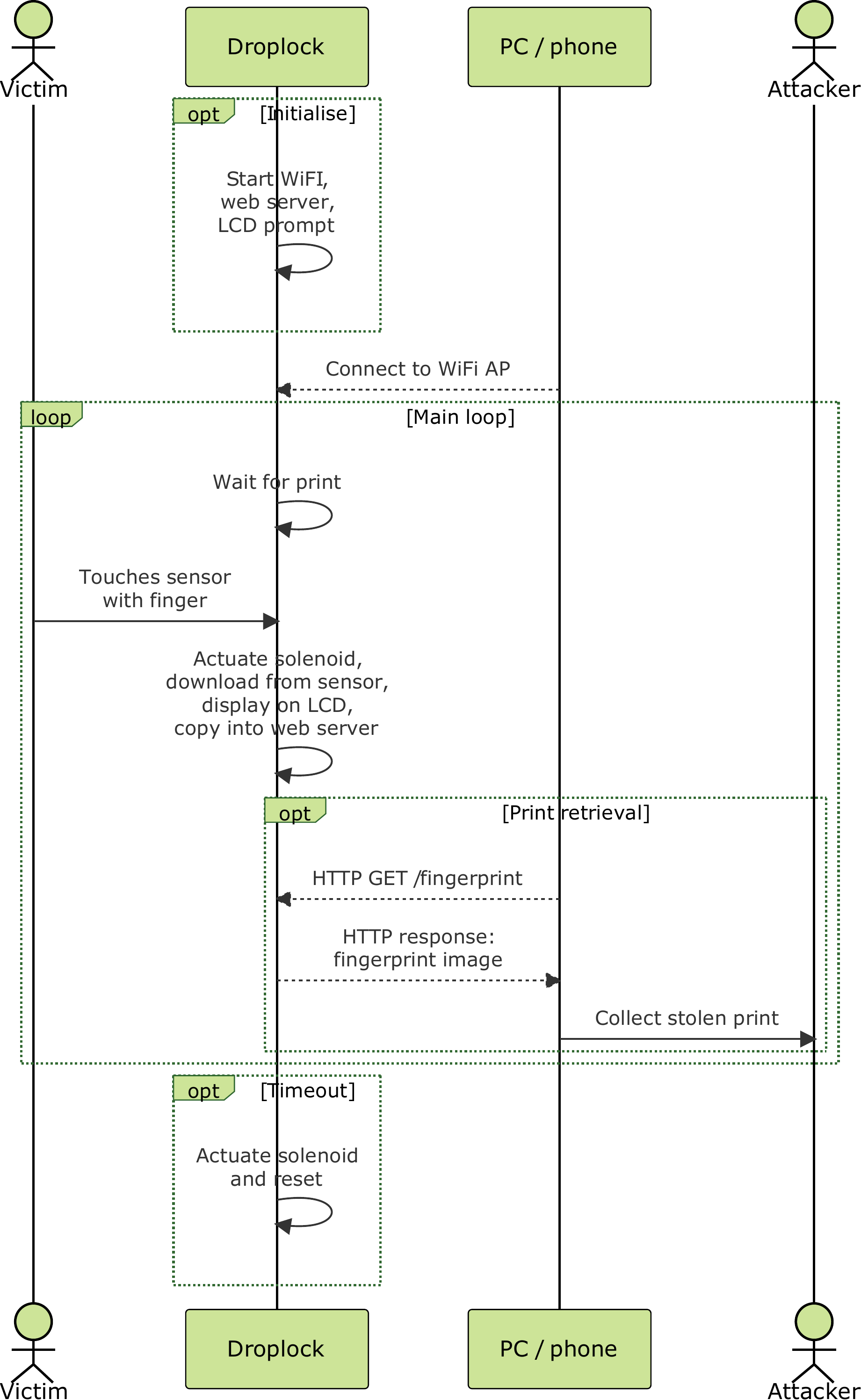}
  \caption{Sequence diagram of \droplock* PoC actions and data flows.}
  \label{fig:pocseq}
\end{figure}

\begin{figure}
  \centering
  \includegraphics[width=.9\linewidth]{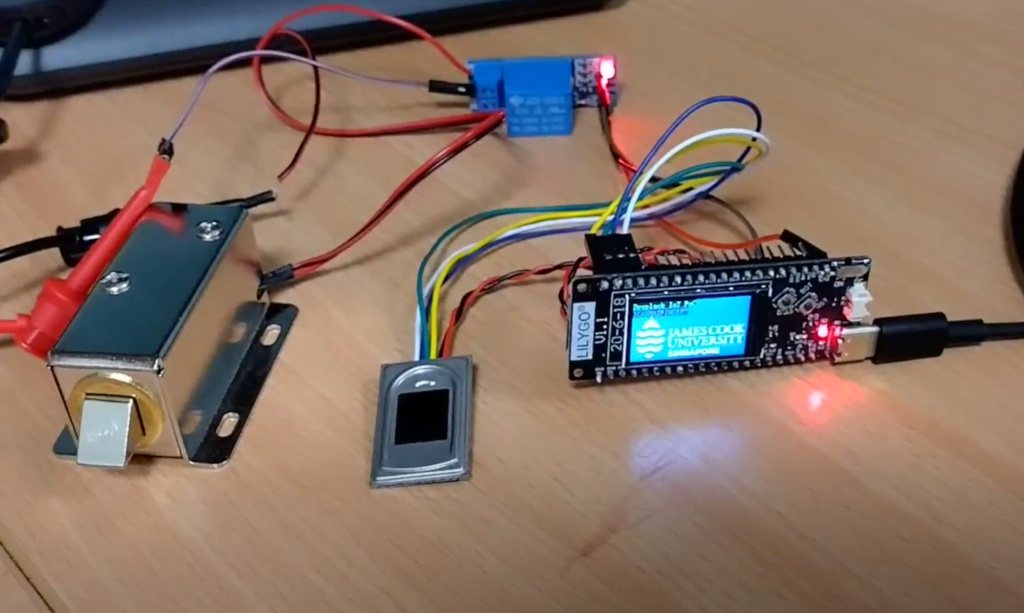}
  \caption{The \droplock* PoC demonstration~\cite{droplock-poc-video}.}
  \label{fig:pocphoto}
\end{figure}

A sequence diagram for the PoC is shown in \cref{fig:pocseq} and explained below:

\begin{enumerate}
  \item The \droplock* device initialises, creates a WiFi AP, web server and prompts for interaction via its LCD screen.
  \item The device polls the fingerprint reader to determine when a print is captured.
  \begin{enumerate}
    \item After 60 seconds of no input, the device will actuate the solenoid to draw attention to itself and then reset.
    \item If a user presents a fingerprint before this deadline, the device will request the reader upload the fingerprint and actuate the solenoid as a reward for interaction.
  \end{enumerate}
  \item Upon acquisition of a fingerprint, the fingerprint reader uploads the image to the micro-controller.
  \item The uploaded fingerprint is displayed on the LCD and the webserver's fingerprint image is updated.
  \begin{enumerate}
    \item For the PoC, a blank image is swapped for a koala fingerprint image once the user's print has been acquired.
    \item In a true implementation, the raw buffer would be re-formatted into web-compatible image.
  \end{enumerate}
  \item The device waits 30 seconds for a user to connect and fetch the image from the embedded web server over WiFi.
  \item The sequence repeats.
\end{enumerate}

The source code of the PoC~\cite{droplock-poc-firmware} and a video demonstration~\cite{droplock-poc-video} have been made available. The physical appearance of the setup is shown in \cref{fig:pocphoto}. After the user presents a fingerprint, it can be uploaded from the reader to the device in approximately two seconds. Fetching the image over WiFi can then be done in a matter of milliseconds, assuming the client has already connected to the device.

Unlike prints left behind through contact with surfaces, which could be wiped away, the captured fingerprint from the \droplock* can be transferred and stored elsewhere immediately, so an errant impression upon the device's fingerprint reader cannot be undone.

Despite demonstrating adequate technical capabilities, the PoC is large and cumbersome compared to a commercial smart padlock or other fingerprint smart lock. An OEM design could be purchased, which an attacker can then modify to include \droplock* capabilities. Such a device could then take a convincing form factor and be far more innocuous in appearance.

\section{Hacking a smart lock to harvest prints}
\label{sec:hacking}

To avoid custom building a viable \droplock*, an attacker may convert a COTS smart padlock into one. Brand recognition or user familiarity with the product can then be leveraged to increase the chance of a successful attack.

We acquired several smart padlock products~---~some branded and others unbranded~---~to find one that could be hacked into a \droplock*. While it is likely that several of the products may be hackable, this paper explores a single successfully hacked lock and looks to future work to survey the larger market.

\subsection{The target lock}
\label{sec:target}

A target lock needs a fingerprint reader and radio communications that can be repurposed to serve the \droplock* goals. Repurposing is done by modifying the firmware on the target device. We envisaged three possible routes to achieving this:

% The selected lock has a fingerprint reader and uses BLE to communicate with a smartphone app for remote unlocking and device management purposes.

% The subject of the attack is the cheapest of the vendor's products. However, it has the same fundamental features of interest to this work: a fingerprint reader and wireless capabilities, in this case in the form of BLE communication with a smartphone app. The intent is to modify or replace the firmware on the device to steal fingerprints and send them via BLE to a custom application on a nearby laptop or smartphone. Prior to physically compromising the device and learning as much about it as possible, three possible approaches were envisaged:

\begin{enumerate}
  \item Find and exploit a vulnerability in the device firmware, such as a buffer overflow, to upload a malicious payload that enables the device to perform \droplock* functions.
  \item Find and exploit a vulnerability in the firmware update process, such as weak or no image validity/signature checking, to program a new \droplock* firmware.
  \item Use a debug interface to directly reprogram the device with a \droplock* firmware and re-assemble it.
\end{enumerate}

In the early stages of assessment, one of the procured locks was found to be vulnerable to option 3, and so with the expanded capabilities of debugger access to the device, we chose to use this method. The device in question features a fingerprint reader and uses BLE to communicate with a smartphone app for remote unlocking and device management.

\subsubsection*{Disclosure of vulnerabilities}

We do not believe that disassembly and malicious reprogramming of smart locks has been considered as a security concern prior to now, except for companies wishing to protect their IP. We also believe that similar products will possess the same issue, and until a broader survey is completed, we have avoided reference to the brand and product where possible. During the course of this work we have reached out to the manufacturer to inform them of the research and any security issues that were discovered and will engage with them should they wish to do so.

\subsection{Physical compromise}
\label{sec:physical}

The lock can be non-destructively dismantled by removing a plastic covering and a specialist screw. The back panel can then be removed and the PCB extracted. These steps are reversible without leaving evidence, provided the padlock is unlocked at the time. This is sufficient if the attacker purchases the target hardware themselves.

\begin{figure*}
  \centering
  \subfloat[Assembled lock]{\includegraphics[height=185px]{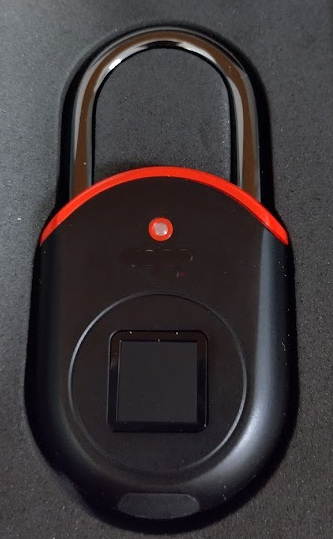}
  \label{fig:lock-front}}
  \subfloat[Lock internals]{\includegraphics[height=185px]{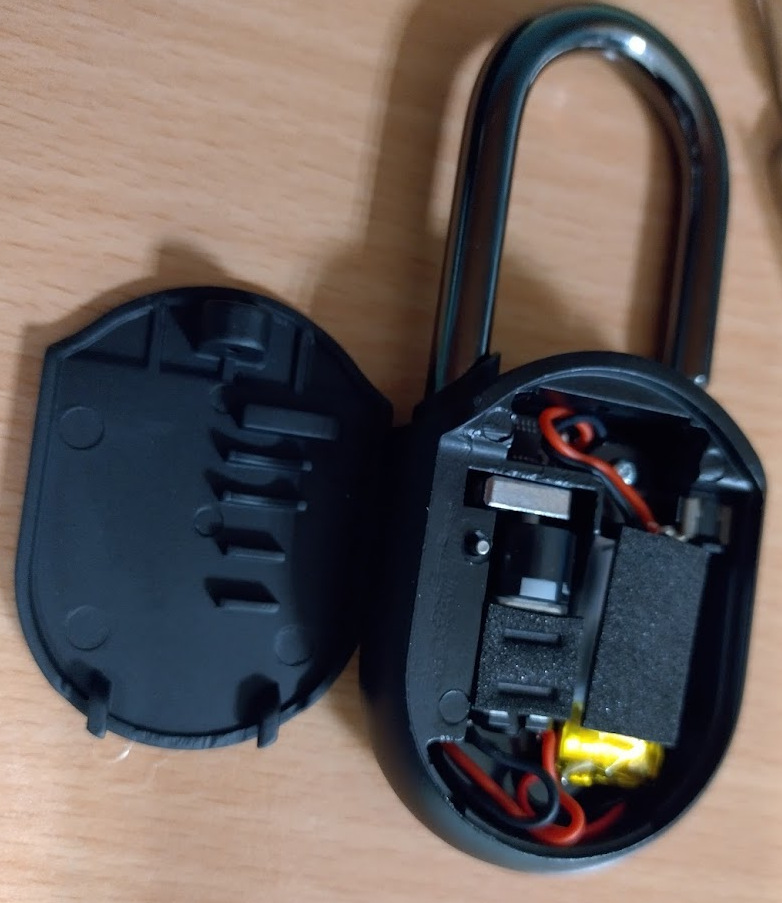}
  \label{fig:lock-open}}
  \subfloat[Lock PCB]{\includegraphics[height=185px]{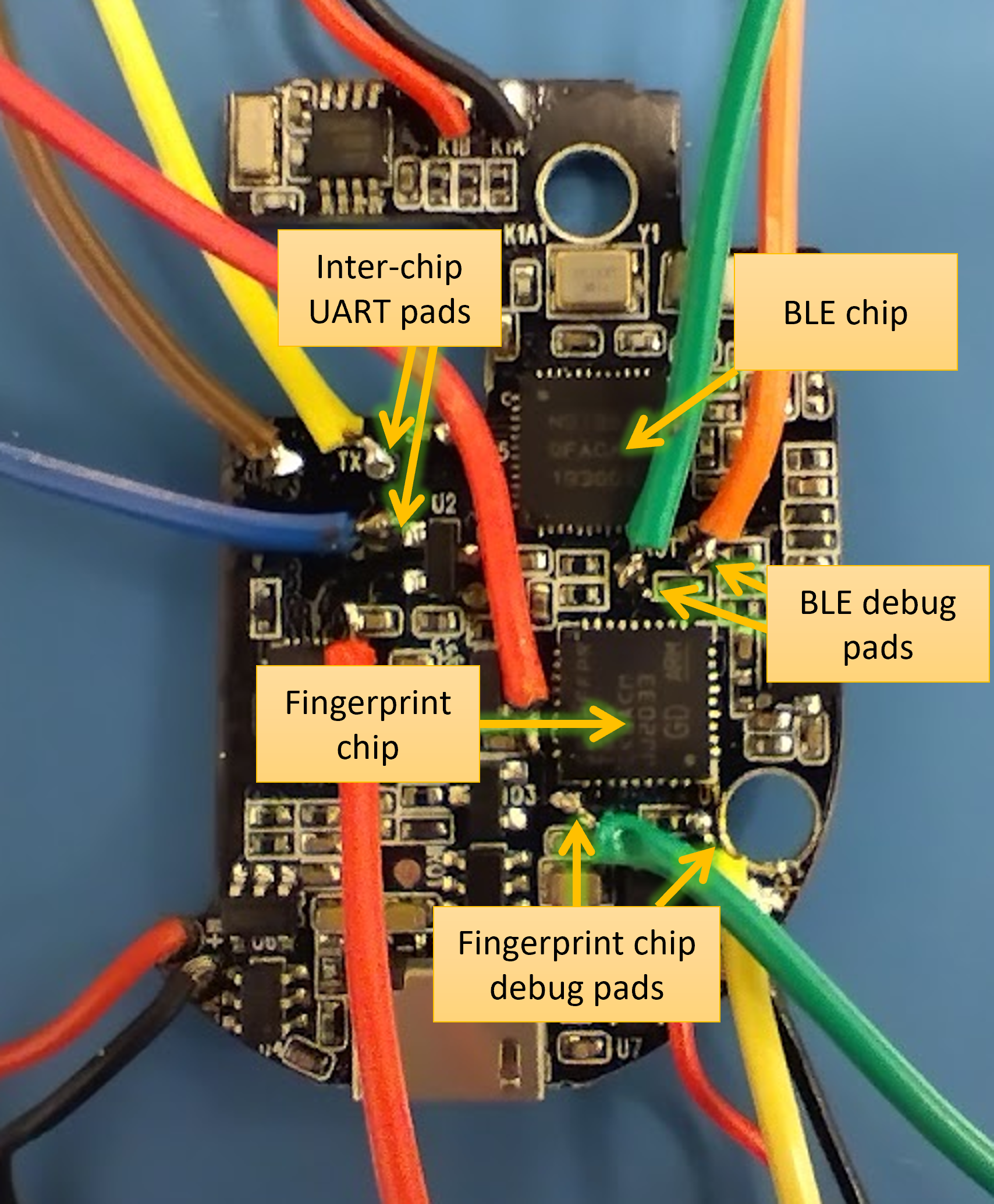}
  \label{fig:lock-pcb-annotated}}
  \caption{Disassembly of the target lock to observe board components and attach cables for debugging, programming and wiretapping.}
  \label{fig:circuitboard}
\end{figure*}

\Cref{fig:circuitboard} shows the target lock progressively disassembled, with \cref{fig:lock-pcb-annotated} showing the PCB with breakout cables soldered to it and annotations of the noteworthy parts. This particular product is a two-chip solution, with the fingerprint sensor controlled by one chip, a GigaDevice STM32 clone, which communicates over UART with a BLE-enabled micro-controller, a Nordic Semiconductor (for brevity, Nordic) nRF51822~\cite{nRF51822}. They are both ARM-based Cortex-M series CPUs. Debug pads for both chips are present on the board. Debug capabilities were found enabled on both chips, with no memory protection. These debug interfaces were used to dump the flash contents for the chips using OpenOCD~\cite{OpenOCD} and an ST-Link v2 debugger/programmer~\cite{STLink}.

Pads are also present on the UART link between the two chips, allowing wiretapping of the communications between them. This, combined with chip-maker's documentation and flash dumps provided enough starting information to develop a \droplock* firmware for the device and reprogram it.

\subsection{Reverse engineering and reprogramming}
\label{sec:reveng}

\begin{figure}
  \centering
  \includegraphics[width=0.9\linewidth]{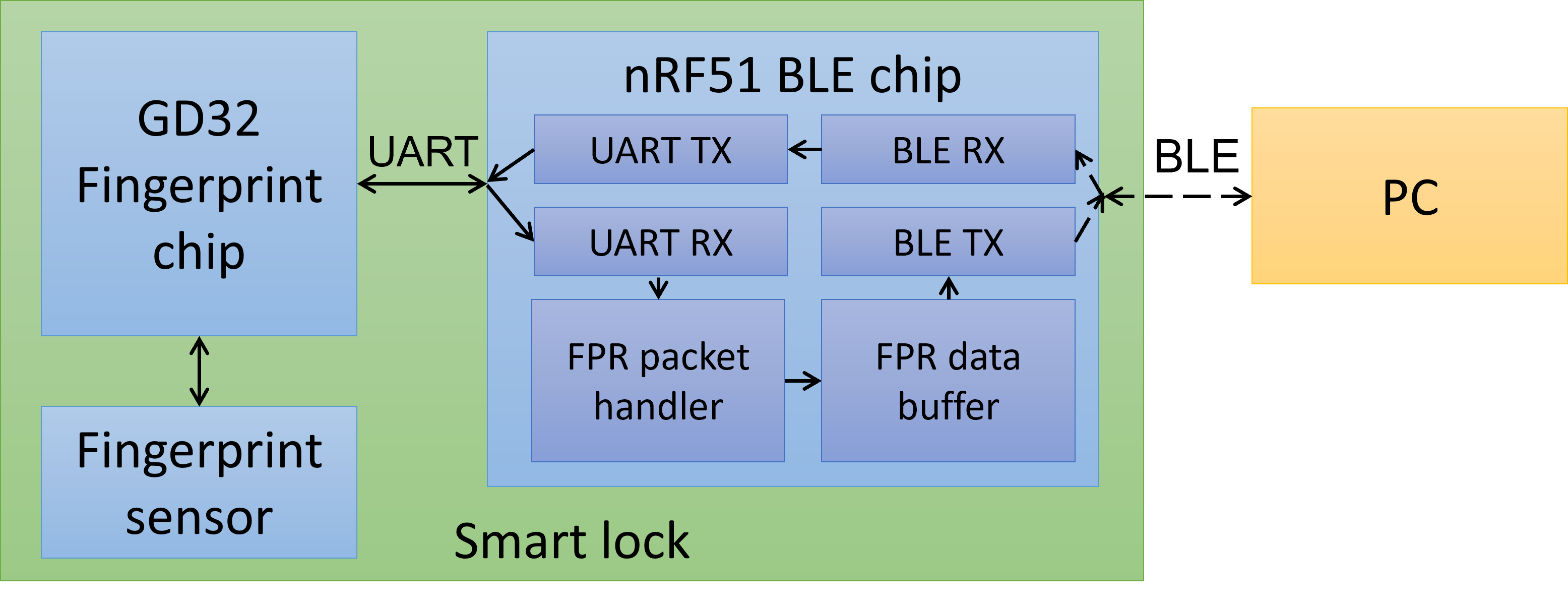}
  \caption{Structure of the hacked device's main components, including data flow through the replacement firmware.}
  \label{fig:hackstructure}
\end{figure}

While some reverse-engineering of the device's firmware was performed using Ghidra, it was determined that an easier modification would be to supply a whole new firmware built within the Nordic's nRF5 SDK~\cite{NordicSDK}. Version 12.3.0 (from 2017) of the SDK was chosen based on the features observed in firmware update packages for the target device, although it is not known exactly which SDK version the original firmware is developed in. \Cref{fig:hackstructure} shows the design of the hacked lock.

A separate micro-controller with two UART interfaces was used to wiretap the communication in both directions between the BLE chip and fingerprint chip on the lock. From these captures it was determined that the device protocol appeared to be the same as the ID809 fingerprint reader~\cite{ID809Product}, which is available to purchase for hobbyist projects and has an open source library~\cite{ID809Source}. This library was used as reference for constructing and interpreting commands between the laptop and fingerprint chip.

OpenOCD was used to observe the configuration and data registers of the device when running the original firmware to determine I/O, clock and other peripheral setup. These were not entirely reverse-engineered, as not all features of the board were needed for the \droplock* firmware to be able to perform an attack.

Nordic's \texttt{ble\_app\_uart} example application, included in the nRF5 SDK, was used as a starting point for the \droplock* firmware. This sample application bridges an nRF51's UART over BLE. The application was modified with a new board support package that adequately fit the lock's hardware design. The UART to BLE bridging of the example application gives a connecting device direct access to the ID809-like control protocol of the fingerprint chip.

One significant challenge was buffering and sending fingerprint data over BLE. The UART on the lock runs at \SI{115200}{bps} baud rate by default, which exceeds the data rate achievable over BLE in the configuration supported by the hardware and SDK version. Additionally the ``SoftDevice'' --- Nordic's proprietary and closed-source portion of the firmware --- only supports 20 bytes of application data per message with this SDK version. This is smaller even than the shortest command and response packets for the fingerprint chip, which are 26 bytes long. Complicating matters further, the fingerprint images are 25,600 bytes which is significantly larger than the free RAM of the device. A quarter resolution version can be requested, but we sought to obtain the maximum quality image from the \SI{508}{dpi} sensor.

These limitations were overcome by implementing a ring buffer and packet handler that would buffer data while interpreting the fingerprint chip's response headers to determine the expected packet lengths, sending fragments of 20 bytes until all expected data was forwarded. These components are depicted as the \emph{FPR packet handler} and \emph{FPR data buffer} in \cref{fig:hackstructure}. On startup, the micro-controller instructs the fingerprint reader to operate at \SI{9600}{bps} rather than \SI{115200}{bps}, which is much closer to the data rate achievable over the BLE layer. With the ring buffer smoothing out any delays in transmission over BLE, this was adequate to obtain a full fingerprint image.

\subsection{Remote compromise}
\label{sec:remote}
%Through additional vulnerability discoveries that we discovered outside of the scope of this work, the firmware update packages for the target lock were also obtained.
Recent versions of Nordic's SDK use PKI and signed firmware images to establish a chain of trust between the update image, its creator and the device. However, the firmware packages for the target lock have no such signatures, only CRC-16 checksums, indicating an older SDK and Device Firmware Update (DFU) process were used, with little protection against modified firmwares.

The lock in question does have some additional defences. These include encryption of the messages between the lock and phone app that involves obtaining a key from the vendor's web API. This was reverse engineered, enabling another script based on the \texttt{bleak} library to communicate with the lock's original firmware and trigger DFU mode. Two ways of achieving this were discovered:

\paragraph*{Authenticated}

The attacker registers the lock and obtains the serial and key from the web API, authenticating with the lock and then activating DFU mode.
\paragraph*{Before registration}

A new, unregistered lock is given arbitrary serial and key values by exploiting security deficiencies in the initialisation process. These chosen values are then used to authenticate and activate DFU mode.  

Locks registered by other users need their keys and serials compromising by some other means in order for the attacker to gain control of the lock. Obtaining these keys and serials would also compromise the security of an unmodified firmware, as the attacker could then unlock any device. This kind of vulnerability has been found previously~\cite{PenTestPartners}, but this method does not work on our target lock and its out-of-the-box firmware version.

\begin{figure}
  \centering
  \includegraphics[width=.67\linewidth]{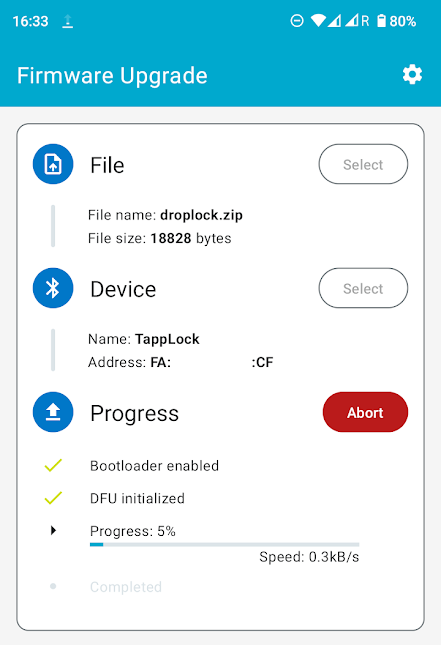}
  \caption{Infecting the lock with \droplock* firmware via DFU over BLE.}
  \label{fig:dfu}
\end{figure}

An DFU package was created using Nordic's \texttt{nfutil} tool (in this case a particularly old version: 0.5.2), based on the same binary that was previously flashed via cables and a debugger/programmer. A DFU app (\cref{fig:dfu}), also available from Nordic, was then used to perform the update once the device had been entered into DFU mode.

This process takes approximately one minute, after which DFU mode can be disabled and the \droplock* firmware becomes usable. At present the \droplock* firmware does not include a DFU activation mechanism, so once a lock has been hacked, it cannot be restored without physical reprogramming. However, this capability could be regained by combining additional components from the nRF5 SDK into the modified \texttt{ble\_app\_uart} code.

This brings the attack closer to an ideal case: the attacker can acquire locks and then reprogram them as \droplock*s before dropping them at target locations and capturing victims' fingerprints. This can even be done without the locks being registered with the manufacturer and without disassembly. Our work has now successfully hacked the target device using both methods 2 and 3 as outlined in \cref{sec:target}.

\subsection{Demonstration}
\label{sec:demonstration}

The source code of the \droplock* firmware for the target lock has been made available~\cite[\textit{Currently private due to ongoing vulnerability disclosure prior to publication}]{droplock-hack-firmware}. A demonstration video has also been produced~\cite{cots-droplock-video}. In the video, the deconstructed lock from \cref{fig:lock-pcb} has had its nRF51822 chip reprogrammed with a custom firmware as described \cref{sec:reveng}. Pressing the button on the device (which when fully assembled is achieved be depressing the lock shackle) wakes the device for approximately one minute.

While awake, the device is discoverable via its MAC address and new name ``IoT Droplock''. A Python script utilising the \texttt{Bleak} BLE library connects to the lock and then issues serial commands to the fingerprint chip via the \droplock* firmware's BLE to UART bridge. The source code for this has also been published~\cite{bleaky-uart}. The host PC commands the lock to wait for a fingerprint impression and then upload it to the PC, after which the image is displayed.

\begin{figure*}
  \centering
  \subfloat[Capture result and victim's thumb.]{\includegraphics[height=80px]{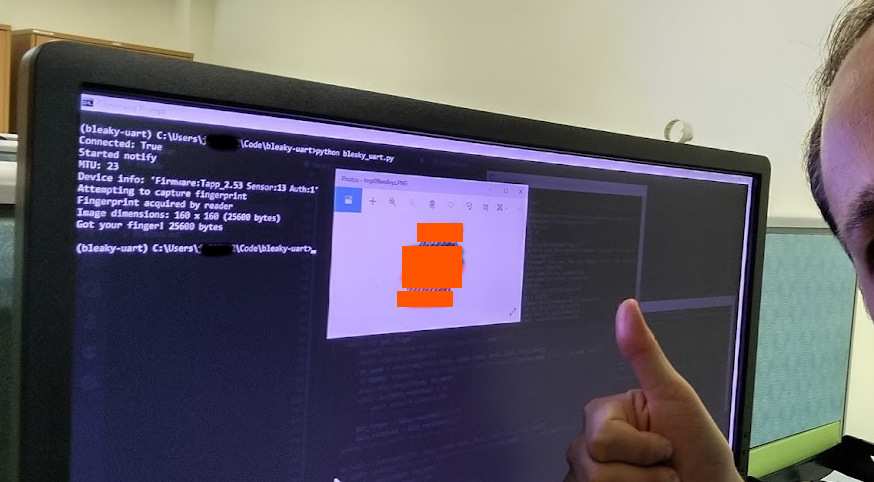}
  \label{fig:droplock-demo-image}}
  \subfloat[Screenshot of capture script output and redacted captured image.]{\includegraphics[height=80px]{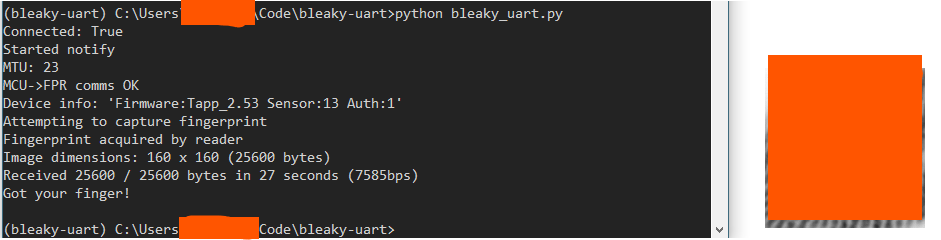}
  \label{fig:droplock-demo-screenshot}}
  \caption{Fingerprint capture from the \droplock* onto a nearby PC with BLE.}
  \label{fig:droplock-demo}
\end{figure*}

The \droplock* sequence is somewhat different to that shown in the PoC from \cref{sec:comppoc}. Firstly, it uses BLE and not WiFi, because of the available technologies on target device. Secondly, fingerprint data is directly streamed to the attacker's PC, rather than buffered into an embedded web server on the device, again due to the communication differences of the target. The attack objective of the lock and the victim's interactions with it remain essentially the same.

Fingerprint uploads take approximately 27 seconds and the data rate after overheads is around \SI{7.5}{kbps}. This is limited by the BLE transmission interval and MTU size. Alternative \droplock* hardware and/or newer SDK versions may be able to use shorter intervals and larger MTUs for faster data rates.

The \droplock* can communicate with a host reliably within a few metres. The attacker's receiving device must be within this range. The Python script could be ported to smartphone, Raspberry Pi or other more easily concealable device. It may also be desirable to introduce a more sophisticated buffering and retransmission mechanism to improve functionality over longer distances.

\section{Related Work}
\label{sec:relatedwork}

Our exploration of prior work is divided into three areas: The propensity for users to interact with dangerous data and devices, the reconstruction of biometrics from sensor data and other security vulnerabilities affecting smart locks.

\subsection{Users interacting with dangerous data and devices}

People have a tendency to unwittingly interact with data or devices that may put them at risk. One such example is the proliferation of Quick Response (QR) codes. These two-dimensional barcode-like images were created by a Japanese automotive company in the 1990s~\cite{QRcode}. They are easy to generate and print or display, and are then scannable by any camera-equipped device with the appropriate software, most commonly a smartphone and app.

QR codes often encode URLs. However, QR codes lack any inherent proof of authenticity, which may allow an attacker to coerce a victim into scanning a QR code that leads them to a malicious website, with objectives including exploit payload delivery and phishing. Efforts have been made to detect malicious URLs in QR codes~\cite{Wahsheh2021}, as well as to allow verification of code authenticity through the embedding of digital signatures into them~\cite{Focardi2019}. Users must utilise an app that integrates such features in order to provide any protection, leaving a majority unprotected and unaware of the risks.

The uses of QR codes include marketing, supply chain tracking, contact tracing, user registrations and many more. They are prolific in society, coupled with the proliferation of smartphones. As embedded IoT devices become equally prolific, user familiarity with them may lead to complacency, especially if good security is not a default posture.

While QR codes can be a visible threat to unaware users, dropped USB sticks and memory cards pose a more tangible threat. A user may find a removable storage device and insert it into a computer to see what is on it, potentially introducing a malicious payload into a system that might otherwise be isolated from attack. It has been shown that users have a tendency to do this~\cite{Tischer2016} and simulations of such attacks are performed by penetration testers during security assessments~\cite{CompTIAPenTest}. USB drops have famously been implicated in high profile incidents involving malware such as Conficker and Stuxnet~\cite{Collins2012}. Removable storage has been used to spread malware for decades, including the first instance of ransomware which was spread by floppy disk in 1989~\cite{Kumari2019}.

If users have a propensity towards scanning QR codes and inserting removable storage without establishing their origin or trustworthiness, then we can infer that users will be similarly vulnerable to picking up and interacting with a \droplock*. In fact, the \droplock* attack may take less effort on the part of the user to be successful. They need only pick up the device and press their finger to the sensor, with no app or physical connection to a computer required.

\subsection{Reconstructing and using stolen biometric data}

Fingerprints are a widely used method of biometric identification. Formerly reserved for high-security scenarios such as door entry systems, they are now used far more widely, including on laptops, smartphones and employee time clocks. In the taxonomy of authentication methods, biometrics are something a person \emph{is}: a physiological aspect to their identity that no other person is likely to have.

However, fingerprints can be reproduced. In criminal forensics for over a century, fingerprint evidence has been used to place individuals at scenes of investigations or to identify people~\cite{Acree1999}. People leave residue on points of contact that can then be lifted by a trained person. Similar techniques can be used for the purpose of stealing a fingerprint.

To impersonate an identity by using a stolen fingerprint, the attacker must be able to reproduce the print in a form that the authentication system will accept. An image of the print may be inadequate, as the biometric sensor may sense the presence of human skin on the sensor's surface, along with other liveness checks such as pulse and temperature. However, 3D printing can be used to produce a scannable surface~\cite{Levalle2020} and liveness checks can be defeated even when modern approaches such as deep neural networks are used~\cite{Fei2020}.

Going some way to defeat the reproduction of fingerprints, some biometric systems will analyse captured fingerprint images and generate a template from it to store that rather than the image itself. This capability can be observed in the sensor used in our target lock~\cite{ID809Source}. Authentication attempts take a scan of a print, generate a new template from it and search the template database for a match. ISO/IEC 19794-2 defines fingerprint standards for interoperability between vendors and systems. However, techniques can be applied to various template formats to reproduce fingerprint images~\cite{Cappelli2007,Arakala2011} to an extent that is sufficient to fool a sensor.

Initially, we had expected to lean on such techniques in order to reproduce fingerprint's from our \droplock*'s template data. However, we have shown that this was not necessary because the fingerprint readers of the devices in \cref{sec:comppoc} and \cref{sec:hacking} provided the original images on request.

\subsection{Smart lock security}

Smart locks are an evolution of electronic or digital locks, adding communication and IoT capabilities. Electronic locks of various kinds have previously been found to have vulnerabilities, such as susceptibility to manipulation of the locking mechanism with magnets~\cite{TooolRing} and easily reproducible key cards for which open source tools and cheap hardware exist~\cite{proxmark3}. Some smart locks have also been found to have physical weaknesses, whereby the spring-loaded mechanism can be overcome with a tap from a hammer~\cite{LPL}.

In addition to physical weaknesses, smart locks may depend on the security of their communication link, both from lock to smartphone and also from smartphone to the manufacturer's cloud service. Previously, models of smart locks and associated systems have lacked adequate encryption on cloud API calls, leaked information via their API through lack of proper authorization~\cite{TapplockFTC} and used easily-reversible key generation methods enabling them to be unlocked by anyone~\cite{PenTestPartners}.

\subsection{Linking everything together}

Combining the prior work we have reviewed above, a full \droplock* attack exploits the following vulnerabilities:

\begin{enumerate}
  \item Users can be coerced into interacting with an unknown device. That device may contain a fingerprint reader.
  \item Fingerprint readers made for or found in IoT devices may provide the full fingerprint image.
  \item These images can be used to reproduce a fingerprint that can fool other biometric security systems.
\end{enumerate}

This process may be more involved than dusting and lifting prints from the surfaces touched by a victim, but is likely to be less obvious in the short-term and may be quicker to perform.

The smart locks may have other weaknesses, physical or digital, that compromise their security in the context of their intended purpose. However, in the case of the \droplock*, its ability to lock something securely is not a concern because it is used only as a lure and is not intended to serve its original function.

\section{Recommendations and future work}
\label{sec:recommendations}

The intent of this work is to highlight the ease with which devices with benign use cases can be repurposed into malicious ones, and that victims may have no way of knowing this. In particular, we suggest that IoT devices, which are small and un-threatening in appearance, should be viewed with the same skepticism and caution as unidentified USB sticks and memory cards. However, while storage devices pose a threat if carelessly inserted into a vulnerable device, a malicious IoT device could be a security threat simply by being handled. The stealthy nature of the data theft is such that the victim will consider the device broken and never suspect that their print has been stolen.

\subsection{Recommendations}

Assuming fingerprints will, like passwords, remain part of authentication processes in the forseeable future, we make the following recommendations for best practices and awareness to mitigate \droplock* risks:

\paragraph{Disable debug}

In addition to a manufacturer being concerned about their product's firmware being retrieved and reversed engineered via debug pads or pins, they should also be concerned about the device being reprogrammed with malicious firmware. Disable debug capabilities on production models to prevent this misuse.

In the case of the nRF51-series chip used in the target lock from \cref{sec:hacking}, it is not possible to fully disable the debug interface, although user configuration registers can disable reading of memory via debug. However, this protection has been circumvented~\cite{nrfsec}, so while it would slow down an attacker, it may not stop them entirely.

\paragraph{Use PKI signed firmware updates}

\Cref{sec:remote} shows that without adequate protections, a \droplock* firmware can be flashed onto a device via an Over The Air (OTA) DFU process. Chipset vendors such as Nordic offer secure signed firmware features in their more recent SDKs, therefore product manufacturers should consider using these wherever possible.

\paragraph{Prevent fingerprint upload}

It should not be possible to retrieve a fingerprint image from the sensor without destructive or otherwise obvious effort. Production devices should at least have this capability disabled. Some portability of data, for example of fingerprint templates, may be acceptable, however these may also still allow reconstruction of fingerprints in some cases and might be best left disabled by default.

\paragraph{Increase user awareness}

A community effort should be made to educate potential IoT device users (i.e. everyone) that an untrustworthy device could affect their security. Aside from fingerprint theft, these devices could be bugged, used as wireless attack vectors or another invisible threat. Beyond simple skepticism towards dropped/found storage devices, we should educate people to be mindful about handling dropped/found IoT devices.

\paragraph{Establishing user trust}

It is impossible to visually determine the authenticity of most IoT devices. While a chain of trust may be present in the device's boot process to ensure that a genuine firmware is running, it may be difficult for a person to verify this without the vendor's custom application. By that time, it may be too late. A standardised method for advertising and verifying origin and integrity, using an IoT device's wireless capabilities would help to mitigate this.

\subsection{Future work}

The most compelling area for further work centres upon studying human behaviour around IoT devices. With the possibility of fingerprint theft demonstrated in this paper, the next step is to determine how susceptible users of various profiles would be to succumbing to such an attack. This information would better motivate the effort required in responding to this threat.

Additionally, formalising a standard for security requirements of any device that possesses a fingerprint scanner that accounts for the \droplock* type of threat would have future security benefits and increase user confidence. So too would researching the development of a standard for close proximity wireless verification of an IoT device's origin and integrity.

Finally, this work examines only one COTS device and demonstrates the \droplock* transformation process on it. A wider examination of a range of COTS products, along with the considerations required to create a convincing device from scratch may help to identify opportunities to improve existing designs as a whole.
\vspace{-1em}
\section{Conclusion}
\label{sec:conclusion}

This paper has shown how a device with biometric and wireless capabilities can be modified to steal fingerprint images. These images can be used with pre-existing techniques to reproduce the victim's prints and then gain access to other systems such as smartphones, bank accounts or offices.

A proof-of-concept of the \droplock* was presented that demonstrated the wireless theft of fingerprint data using easily obtainable hardware. Then, a COTS device was disassembled and reprogrammed with a malicious firmware. The process was physically reversible, meaning the hacked device would look no different once re-assembled. A method of reprogramming the target lock wirelessly and with no disassembly, using DFU over BLE, was also demonstrated. This culminates in an IoT device that may carry trusted or familiar branding, but that has been secretly transformed into a fingerprint harvester.

The security implications of fingerprint theft were explored and recommendations made for improving the security posture of both IoT devices and the public with respect to this vulnerability. Further work to study human behaviour towards dropped/found IoT devices, and how to establish trust beyond visual inspection were proposed.

\vspace{-0.5em}
\section*{Acknowledgment}

The author wishes to thank Denise Dillon and Robert
Bhurke for their advice, and Belinda Lee and Kevin Wang for
administrative support. Finally, thank you to the reviewers for
their feedback.

%%% ===============================================================================
%%% Bibliography
%%% ===============================================================================

% In the bibliography, use \texttt{\textbackslash textsuperscript} for \enquote{st}, \enquote{nd}, \ldots:
% E.g., \enquote{The 2\textsuperscript{nd} conference on examples}.
% When you use \href{https://www.jabref.org}{JabRef}, you can use the clean up command to achieve that.
% See \url{https://help.jabref.org/en/CleanupEntries} for an overview of the cleanup functionality.


% Enable to reduce spacing between bibitems (source: https://tex.stackexchange.com/a/25774)
% \def\IEEEbibitemsep{0pt plus .5pt}

\vspace{-1em}
\bibliographystyle{IEEEtranN} % IEEEtranN is the natbib compatible bst file
\bibliography{paper}

% Enfore empty line after bibliography
\ \\
All links were last followed on 29th April 2022.

\end{document}